\documentstyle[preprint,aps,prd]{revtex}

\def\slash#1{\setbox0=\hbox{$#1$}           
   \dimen0=\wd0                                 
   \setbox1=\hbox{/} \dimen1=\wd1               
   \ifdim\dimen0>\dimen1                        
      \rlap{\hbox to \dimen0{\hfil/\hfil}}      
      #1                                        
   \else                                        
      \rlap{\hbox to \dimen1{\hfil$#1$\hfil}}   
      /                                         
   \fi}

\def\pp{
\setbox0=\hbox{$\frac{d\omega}{\pi}$}         
\dimen0=\wd0                                  
\rlap{\hbox to \dimen0{\hfil\hglue 0.7mm $-$\hfil}}
\int }                                        

\def\leq{\raisebox{0.5ex}{\hbox{$<$}}         
\kern -0.7em\raisebox{-0.4ex}{\hbox{$\sim$}}} 

\begin{document}

\title{ {\bf Self-consistent approximations in relativistic plasmas:\\
       Quasiparticle analysis of the thermodynamic properties}\footnote 
    {\baselineskip=12pt It is a pleasure to dedicate this paper to Leo
Kadanoff on his sixtieth birthday.  Immediately after Leo and I received our
Ph.D.'s from Harvard in 1960, we began working together at the Institute for
Theoretical Physics in Copenhagen (now the Niels Bohr Institute) on the
problem of how to construct self-consistent approximations to two-particle
propagators that preserved the basic conservation laws.  Our solution to the
problem was written up in the paper {\it Conservation laws and correlation
functions} \cite{BaymKada61}, and the concept generalized to deriving
self-consistent approximations from a functional, $\Phi$, of the single
particle Green's function in Ref.~\cite{Phiconser2}.  I hope that Leo enjoys
revisiting these early ideas, which may prove useful in understanding modern
problems of the thermodynamics and transport properties of systems with
long-ranged gauge fields.  -- Gordon Baym}}
\author{Beno\^\i t Vanderheyden and Gordon Baym  \\[0.7em]
 Department of Physics, University of Illinois at Urbana-Champaign,\\
       1110 W. Green St., Urbana, IL 61801, U.S.A.}

\date{\today}
\maketitle

\begin{abstract}
    We generalize the concept of conserving, $\Phi$-derivable, approximations
to relativistic field theories.  Treating the interaction field as a dynamical
degree of freedom, we derive the thermodynamic potential in terms of fully
dressed propagators, an approach which allows us to resolve the entropy of a
relativistic plasma into contributions from its interacting elementary
excitations.  We illustrate the derivation for a hot relativistic system
governed by electromagnetic interactions.
\end{abstract}

\newpage

\section{Introduction}

    The motivations for studying relativistic plasmas are numerous.
Nuclear matter at very high densities is expected, from the asymptotic
freedom of QCD, to be in the form of a plasma of deconfined quarks and
gluons~\cite{CollPer75}.  Such plasmas were present in the early
universe and may exist in the cores of neutron stars; current
experiments aim to produce and study them in collisions of
ultrarelativistic nuclei~\cite{QM96}.  Electromagnetic plasmas are
also of interest as Abelian models of quark-gluon plasmas. In addition, 
problems of relativistic plasmas are closely related to current issues in
condensed matter theory.  In gauge fields models of high $T_c$
superconductors and of the fractional quantum Hall
effect~\cite{HalpLeeRead95}, the magnetic component of the interaction
plays a fundamental role; this component is suppressed in normal
metals by a factor $(v_F/c)^2$ (where $v_F$ is the Fermi velocity),
but is important in both strongly correlated systems and relativistic
gases.

    The thermodynamics and quasiparticle modes of relativistic plasmas
have been the subject of much earlier work.  In order to understand
the equation of state of interacting relativistic plasmas, the
thermodynamic potential has been calculated up to the first six orders
of perturbation theory in the coupling
constant~\cite{AkhiPele60,FreeMcLe77,thermopot}. Studies of the plasma
microscopic properties have also revealed the existence of
well-defined quasiparticle excitations and collective
modes~\cite{plasmons,fermions}.  In particular, the fermionic spectrum
has certain remarkable features not encountered in non-relativistic
plasmas: the spectrum has a gap at zero momentum, and splits into two
branches at small momenta~\cite{fermions}.

    Many physical quantities of relativistic plasmas are infrared
divergent when evaluated in perturbation theory, as a consequence of
the lack of static screening of magnetic interactions.  Taking into
account dynamical screening of such interactions at long wavelengths
(the anomalous skin effect \cite{reuter}), equivalent to including
effects of Landau damping in the photon polarization operator,
eliminates the divergences in transport coefficients~\cite{BaymMoni}
and in the rate of energy loss of fast particles~\cite{BraaThom}.
Polarization effects also modify the interaction between low energy
quasiparticle modes.  The ``Hard Thermal Loop'' expansion scheme,
proposed by Braaten and Pisarski\cite{HTL}, handles the effects of the
medium diagrammatically by introducing vertex and self-energy
corrections at small four-momentum.  This description can also be cast
in terms of kinetic equations~\cite{kinetic}.

    Even after taking into account corrections from medium effects,
difficulties still remain.  Due to a lack of static screening in the
magnetic component of gauge interactions, the fermion quasiparticle
damping rate at finite temperatures is still divergent in perturbation
theory.  Blaizot and Iancu have shown that \cite{irfermion} in QED at
finite temperatures, including multiple scattering \`a la
Bloch-Nordsieck in the vacuum leads to well-defined, divergence-free,
quasiparticle modes.  This structure of the fermion propagator also
appears in gauge field models of the fractional quantum Hall effect
\cite{MuccLee}.  However, it is not clear whether the infrared
structure in the fermion spectrum actually affects physical
observables such as transport coefficients and the specific
heat~\cite{Cvgauge}.  Even quantities that are well-controlled order
by order in perturbation theory can cause difficulties.  The
(asymptotic) expansion of the thermodynamical potential converges very
slowly, which has lead to suggestions to reorganize the perturbation
expansion in terms of dressed fermion states rather than those of a
free interacting gas~\cite{screepert}.

    Our aim here is to give a general framework for analyzing the
effects of different characteristics of the fermion spectrum, the
presence of a gap and the infrared structure, on thermodynamic
quantities.  Our analysis is based on $\Phi$-derivable conserving
approximations, first introduced~\cite{Phiconser2,Phiconser1} in the
context of quantum transport theories.  Among various applications,
their use in the study of liquid $^3$He has allowed one to interpret
the $T^3 \log T$ term in the low temperature specific heat in terms of
repeated scattering of particle-hole pairs~\cite{He3,Riedel}.  The
technique has been extended to Bose condensed systems in Refs.
\cite{Gotze}.  A generalization of the conserving approximation
techniques to relativistic plasmas at zero temperature provided a
controlled expansion of the ground state energy of electromagnetic and
quark plasmas up to order $g^4$~\cite{FreeMcLe77}.  In this latter
work, the free energy is given as a functional of the fully dressed
fermion and boson propagators as well as fully dressed vertices.  In
our present analysis, we keep track of both fermion and boson
elementary excitations and treat both the matter and interaction
fields as dynamical quantities.  This allows us to decompose the
entropy in the form
\begin{eqnarray}
s=\sum_{\sigma, {\bf p}} \int {d\omega_p\over 2\pi} \sigma_f(\omega_p)
A_s(\omega_p,p)+\sum_{{\rm pol},{\bf q}} \int {d\omega_q\over 2 \pi}
\sigma_b(\omega_q) B_s(\omega_q,q)+s',
\end{eqnarray}
where the entropy densities of free fermion and boson gases,
$\sigma_f= -f \log f -(1-f)\log(1-f)$ and $\sigma_b= -n\log
n+(1+n)\log(1+n)$ are here weighted by the spectral densities $A_s$
and $B_s$ of the interacting system.

    In this paper, we develop $\Phi$-derivable approximations for a
hot relativistic QED plasma, a gas of electrons and positrons in a
thermal bath of photons at temperature $T$, where $T\gg m$, although
we expect the approach to be valid also for non-Abelian theories
including QCD.  For simplicity, we set the electron mass, $m$, to
zero.  We then illustrate the technique by computing the entropy
within the one-loop approximation, and comment briefly on the general
structure of the entropy spectral densities $A_s$ and $B_s$.

\section{The thermodynamic potential}

    We first review $\Phi$-derivable approximations in
non-relativistic field theories~\cite{Phiconser2}.  The
thermodynamical potential of a non-relativistic Fermi system can be
written as a functional of the fully dressed fermion propagator $G$
as,
\begin{eqnarray}
\beta \Omega &=& \Phi[G]+{\rm tr\,} \Sigma G - {\rm tr\,} \log(-G).
\label{omega:nr}
\end{eqnarray}
Here $\beta=1/T$ and the symbol ${\rm tr\,}$ denotes a trace over
spins and coordinates.  In four-dimensional notation,
\begin{eqnarray}
{\rm tr\,} X&\equiv&\sum_{\sigma,\sigma'} \int_0^{-i \beta} d1 X(1,1^{+}),
\end{eqnarray}
where $1$ is shorthand for the coordinates $(t_1,{\bf r_1})$, while
$1^{+}$ indicates that the time variable in the second argument of $X$
is $t_1+i 0^{+}$.  The functional $\Phi[G]$ is defined as the sum of
all skeleton graphs contributing to the diagrammatic expansion of the
potential $\Omega$, where all lines are fully dressed propagators,
$G$, instead of bare ones, $G_0$.  In contrast with the familiar
diagrammatic expansion of a scattering diagram in the vacuum, the
expansion for $\Phi[G]$ contains a symmetry factor $1/n$ for each
diagram with $n$ fermion lines.  Upon variation of $G$, the functional
$\Phi[G]$ satisfies the relation
\begin{eqnarray}
\delta \Phi[G]={\rm tr\,} \Sigma \delta G,
\label{deltaphi:nr}
\end{eqnarray}
where $\Sigma$ is the electron self-energy.
Pictorially, this relation means that removing one of the $n$ fermion
lines in a given $\Phi$-diagram produces a self-energy diagram with
$n-1$ fermion lines.  Since there are $n$ ways of removing a line, no
symmetry factor appears on the right side of Eq.~(\ref{deltaphi:nr}).
The self-energy $\Sigma$ is here a functional of the fully dressed
propagator $G$, and it satisfies Dyson's equation
$G^{-1}=G^{-1}_0-\Sigma$, where the bare propagator $G_0$ is given by
$G_0^{-1}=\delta_{\sigma,\sigma'} (\omega_n-p^2/2 m)$ for a
non-relativistic system of fermions.

    The principle behind conserving approximations is the following:
one selects a particular subset of diagrams for the functional,
$\Phi_a[G]$, from which one deduces an approximate self-energy
functional $\Sigma_a\equiv \delta \Phi_a[G]/ \delta G$.  Dyson's
equation provides then a self-consistent relation for the approximate
propagator $G_a$.  As shown in \cite{Phiconser2}, the approximation
for $G$ leads to current densities and an energy-momentum tensor that
obey the continuity equations expressing charge, particle number,
energy and momentum conservation.  The basic ingredient of the
approximation scheme of Eq.~(\ref{omega:nr}) which enforces
conservation laws is the stationarity property of $\Omega$,
$
\delta \Omega=0,
$
upon a variation of $G$ that keeps $G_0$ constant, as can easily be
checked from Eqs.~(\ref{omega:nr}) and (\ref{deltaphi:nr}).

    We now generalize this approach to relativistic field theories,
illustrating the method for electromagnetism, where the Lagrangian is
\begin{eqnarray}
\cal L &=& \bar \psi (i \slash{\partial}- e \slash{A}) \psi - {1\over 4}
F^{\mu\nu} F_{\mu\nu}.
\label{lagrangian:QED}
\end{eqnarray}
In QED, the free energy $\beta\,\Omega$ is given in terms of the fully
dressed electron and photon propagators $G$ and $D$, by
\begin{eqnarray}
\beta\,\Omega&=&\Phi[G,D]-{\rm Tr\,} \Sigma G+
{\rm Tr\,} \log(-\gamma^0 G)+{1\over 2} {\rm Tr\,}\Pi D
-{1\over 2}{\rm Tr\,}\log(-D)\equiv W[G,D].
\label{W:QED}
\end{eqnarray}
Here, the functional $\Phi[G,D]$ is the sum of all the skeleton
diagrams of the thermodynamic potential, expressed in terms of fully
dressed $G$ and $D$ instead of the bare electron and photon
propagators, $G_0$ and $D_0$.  Under a simultaneous variation of the
propagators $G$ and $D$,
\begin{eqnarray}
\delta \Phi[G,D]&=&{\rm Tr\,} \Sigma \delta G-{1\over 2} {\rm Tr\,}\Pi 
\delta D,
\label{deltaphi:QED}
\end{eqnarray}
as one easily sees by removing an electron line or a photon line in a given
diagram contributing to $\Phi$.  The electron self-energy $\Sigma[G,D]$ and
the polarization operator $\Pi[G,D]$ are here functionals of $G$ and $D$ and
satisfy Dyson's equations
\begin{eqnarray}
G^{-1}&=&G_0^{-1}-\Sigma, \\
D^{-1}&=&D_0^{-1}-\Pi.
\label{Dyson}
\end{eqnarray}

    To prove that the functional $W[G,D]$ in Eq.~(\ref{W:QED}) is identical to
the thermodynamic potential $\beta \Omega$ of an electromagnetic plasma, we
scale the coupling by a parameter $\lambda$ in the interaction Lagrangian,
\begin{eqnarray}
\cal L_{\rm int} &=& -\lambda e \bar \psi \slash{A}\psi.
\label{lagrangian,lambda:QED}
\end{eqnarray}
We work in Coulomb gauge.  In the non-interacting limit, $\lambda \to 0$,
both $\Sigma$ and $\Pi$ vanish, and we recover the result for a gas of free
electrons and photons
\begin{eqnarray}
\beta \Omega_0 &=& W[G_0,D_0] = {\rm Tr\,} \log (-\gamma_0 G_0)
 - {\rm Tr\,}\log (-D_{0,T}),
\label{omega0:QED}
\end{eqnarray}
where only transverse polarizations (subscript $T$) appear in the
photon contribution~\cite{remark}.  To prove that
$\beta\,\Omega=W[G,D]$ it suffices to show that
\begin{eqnarray}
\beta {\partial \Omega_\lambda \over \partial \lambda}&=&{\partial W\over 
\partial \lambda}[G_\lambda,D_\lambda],
\label{whatweshow}
\end{eqnarray}
for arbitrary $\lambda$.

    Consider first the quantity $\beta\partial\Omega_\lambda/\partial
\lambda$.  From the definition of the thermodynamical potential,
\begin{eqnarray}
\Omega_\lambda &\equiv&-T \log
{\rm tr\,}[\exp\left(-\beta \left(H_\lambda-\mu Q\right)\right)]
\label{partialomega:QED}
\end{eqnarray}
(where $Q$ is the charge operator, the Hamiltonian is $H_\lambda= H_0-
\int d^3 {x} \,\cal L_{\rm int}$, and ${\rm tr\,}$ denotes a sum over
all the quantum states of the system,) we find
\begin{eqnarray}
\beta{\partial \Omega_\lambda \over \partial \lambda}&=&{\beta\over\lambda}
\langle{H_{\rm int}}\rangle=-{\beta\over\lambda} \int d^3 {x} \,
\langle{\cal L_{\rm int}}\rangle  .
\label{omegaHint}
\end{eqnarray}
Here, $\langle{X}\rangle$ denotes the thermal average ${\rm
tr\,}[\exp(-\beta(H_\lambda-\mu Q))\,X ]/{\rm
tr\,}[\exp(-\beta(H_\lambda-\mu Q))]$.  We evaluate next ${\partial
W[G_\lambda,D_\lambda]/\partial \lambda}$.  A given diagram
contributing to $\Phi[G_\lambda,D_\lambda]$ depends on $\lambda$
implicitly through the electron and photon propagators, $G_\lambda$
and $D_\lambda$, but also explicitly through the coupling $\lambda e$,
Eq.~(\ref{lagrangian,lambda:QED}).  However, the terms arising from
the dependence on $\lambda e$ through $G_\lambda$ and $D_\lambda$
cancel in the partial derivative, since $W[G,D]$ is stationary under a
variation of $G$ and $D$.  From the definition of $\Phi[G,D]$, the
variation of $W$ is
\begin{eqnarray}
\delta W[G,D]&=&{\rm Tr\,} \Sigma \delta G-{1\over 2}{\rm Tr\,}\Pi 
\delta D-{\rm Tr\,} \delta(\Sigma G)+{1\over 2}{\rm Tr\,}\delta (\Pi D) 
+ {\rm Tr\,} \delta(\log(-\gamma_0 G))\nonumber \\
&&- {1\over 2}{\rm Tr\,}\delta(\log(-D)).
\end{eqnarray}
Using $\delta \log(-\gamma_0 G)= -G \delta G^{-1}$ and Dyson's
equation, $\delta G^{-1}=-\delta \Sigma$, as well as similar relations
for the photon propagators, we find that
\begin{eqnarray}
\delta W[G,D]=0,
\label{deW}
\end{eqnarray}
under a variation that does not affect $G_0$, $D_0$, or the coupling
constant, $\lambda e$.  Therefore, in the derivative of $W$ with respect to
$\lambda$ only the explicit dependence in the coupling constant contributes,
and we find
\begin{eqnarray}
{\partial W \over \partial \lambda}[G_\lambda,D_\lambda]&=&\left.{\partial
\Phi\over \partial \lambda}[G_\lambda,D_\lambda]\right|_{G,D},
\label{derivW:QED}
\end{eqnarray}
where the derivative on the right side is taken at fixed $G$ and $D$.

    Next, to compute the right side of Eq.~(\ref{derivW:QED}), we use the
invariance properties of $\Phi$.  As each vertex of a given $\Phi$-diagram is
connected to two electron lines and one photon line, the functional $\Phi$
remains constant under the scaling transformations
\begin{eqnarray}
\Phi[G_{\lambda},D_{\lambda}]&=&\Phi[s^{-f} G_{\lambda},s^b
D_{\lambda};s\lambda ], \quad f=1+b/2, \label{scale:QED} \\
\Phi[G_{\lambda},D_{\lambda}]&=&\Phi[s^{-\bar f} G_{\lambda},s^{\bar b}
D_{\lambda};\lambda], \quad \bar f=\bar b/2, \label{scale2:QED}
\end{eqnarray}
where the third arguments of $\Phi$ on the right sides are the scaling
factors of the coupling constant.  Taking the derivative of
Eq.~(\ref{scale:QED}) with respect to $s$, one finds
\begin{eqnarray}
\lambda \left. {\partial\Phi\over\partial (s \lambda)}\right|_{G,D}
- f s^{-f-1} {\rm Tr\,}\,{\delta\Phi\over\delta G}\,G_\lambda
+b s^{b-1}{\rm Tr\,}{\delta\Phi\over\delta D}\,D_\lambda=0,
\label{derivscale:QED}
\end{eqnarray}
which, for $s=1$, gives
\begin{eqnarray}
\lambda \left.{\partial\Phi\over\partial \lambda}\right|_{G,D}
= f\,{\rm Tr\,} \Sigma_\lambda G_\lambda+{b\over 2}\, 
{\rm Tr\,}\Pi_\lambda D_\lambda=f\,{\rm Tr\,} \Sigma_\lambda G_\lambda
+(f-1) \, {\rm Tr\,}\Pi_\lambda D_\lambda.
\label{var1:QED}
\end{eqnarray}
Similarly, the derivative of Eq.~(\ref{scale2:QED}) with respect to $s$ gives
\begin{eqnarray}
{\rm Tr\,} \Sigma_\lambda G_\lambda + {\rm Tr\,}\Pi_\lambda D_\lambda=0.
\label{SGPiD:QED}
\end{eqnarray}
Therefore, Eqs.~(\ref{SGPiD:QED}),~(\ref{var1:QED}) and~(\ref{derivW:QED}) give
\begin{eqnarray}
{\partial W \over \partial \lambda}[G_\lambda,D_\lambda]&=&{1\over\lambda}
{\rm Tr\,} \Sigma_\lambda G_\lambda.
\label{WvsSG:QED}
\end{eqnarray}

    To complete the proof, we relate ${\partial {\rm Tr\,}
\Sigma_\lambda\,\Sigma_\lambda/\partial \lambda}$ to $\langle{H_{\rm
int}}\rangle$, starting with the equation of motion for the
time-ordered fermion propagator $G_\lambda(1,1')\equiv -i \langle{T
\psi(1) \bar \psi(1')}\rangle$, which, from the Lagrangian,
Eq.~(\ref{lagrangian:QED}), is
\begin{eqnarray}
i\slash{\partial}_1 G_\lambda(1,1')+ i \lambda e \langle{T \slash{A}(1)
\psi(1)\bar\psi(1')}\rangle
&=& \delta(1-1').
\label{eqmotionG}
\end{eqnarray}
Comparing to Dyson's equation in real space,
\begin{eqnarray}
\int d\bar 1\, G_0^{-1}(1,\bar 1) G_\lambda(\bar 1,1')-\int d\bar 1
\,\Sigma_\lambda(1,\bar 1) G_\lambda(\bar 1,1')&=&\delta(1-1'),
\end{eqnarray}
we identify
\begin{eqnarray}
{\rm Tr\,} \Sigma_\lambda G_\lambda&\equiv&\int d 1\int d\bar 1\,
\Sigma_\lambda(1,\bar 1)\,G_\lambda(\bar 1,1^{+})= 
i \lambda e \int_0^{-i\beta} d 1 \langle{\bar\psi(1)\slash{A}(1)\psi(1)}\rangle
\nonumber \\
&=&\beta \lambda e \int d{\bf r_1} \langle{\bar\psi({\bf r}_1,0)
\slash{A}({\bf r}_1,0)\psi({\bf r}_1,0)}\rangle
= \beta \langle{H_{\rm int}}\rangle.
\end{eqnarray}
Therefore,
\begin{eqnarray}
\beta{\partial W\over \partial \lambda}={1\over\lambda}{\rm Tr\,} 
\Sigma_\lambda\,G_\lambda
={\beta\over\lambda}\langle{H_{\rm int}}\rangle=\beta{\partial 
\Omega\over\partial \lambda},
\end{eqnarray}
which completes the proof.

\section{The entropy}

    We now turn to the derivation of the entropy of the plasma in terms of the
fully dressed propagators $G$ and $D$, which is given by the derivative
\begin{eqnarray}
S&=&-\left.{\partial \Omega \over \partial T}\right|_{V,\,\mu}
=-\left.{\partial \over \partial T} (T W[G,D])\right|_{V,\,\mu},
\label{seqdodt}
\end{eqnarray}
at constant volume $V$ and chemical potential $\mu$.  We work in
frequency-momentum space, where the electron and boson propagators have the
spectral representations
\begin{eqnarray}
G({\omega_n},{\bf p})&=&\int
\frac{d\omega_p}{2\pi}\,\frac{A(\omega_p,{\bf p})}{\omega_n-\omega_p},
\label{Gspec}\\
D_{T}({\omega_n},{\bf q})&=&\int
\frac{d\omega_q}{2\pi}\,\frac{B_{T}(\omega_q,{\bf
q})}{\omega_n-\omega_q},\label{Dtspec} \\
D_{L}({\omega_n},{\bf q})&=&{1\over q^2}+\int
\frac{d\omega_q}{2\pi}\,\frac{B_{L}(\omega_q,{\bf
q})}{\omega_n-\omega_q},
\label{Dlspec}
\end{eqnarray}
where the subscripts $L$ and $T$ denote longitudinal and transverse
polarizations.  The Matsubara frequencies are $\omega_n=i \pi (2 n+1) T+\mu$
for electrons and $\omega_n = 2 i \pi n T$ for photons.  The functional
$W[G,D]$ depends therefore on the temperature through the spectral functions
$A$ and $B$, through the complex frequencies $\omega_n$, as well as through
factors $T$ at the vertices~\cite{BaymKada62}.  Because $W[G,D]$ is stationary
under a simultaneous variation of the propagators $G$ and $D$, Eq.~(\ref{deW}),
it is also stationary under variations of $A$ and $B$ that ignore the other
temperature dependences.  Thus, the contribution to the entropy
coming from the temperature derivatives of the spectral densities alone
vanishes.  In the following, we therefore evaluate every derivative with
respect to the temperature at constant $A$ and $B$.

    We start from the expression of the thermodynamical potential
$\beta\,\Omega=W[G,D]$, Eq.~(\ref{W:QED}).  Evaluating the frequency sum by
standard contour integration techniques~\cite{BaymKada62}, we find
\begin{eqnarray}
\Omega&=&T \Phi[G,D]+\sum_{\bf p}\int {d\omega_p\over  \pi}
f(\omega_p)\,{\rm tr\,} {\rm Im}\, \left[\Sigma G + \log(-\gamma_0 G^{-1})
\right] \nonumber \\
&+& \sum_{\bf q}\pp {d\omega_q\over 2 \pi}
\,n(\omega_q)\,  \sum_{l=L,T} g_l\, {\rm Im} \left[\Pi_l\, D_l
+ \log(-D_l^{-1})\right],
\label{omegafreq}
\end{eqnarray}
where $f(\omega_p)=(\exp(\beta (\omega_p-\mu))+1)^{-1}$, and
$n(\omega_q)=(\exp(\beta \omega_q)-1)^{-1}$ are the Fermi and Bose occupation
factors, ${\rm tr\,}$ is a trace over spinor indices, the degeneracy factors 
$g_l$ are $g_L=1$ for longitudinal modes and $g_T=2$ for transverse ones.  
Here, ${\rm Im}\,[F]$, means ${\rm Im}\,[F(\omega+i 0^{+})]$.  The integration 
over $\omega_q$ is a principal value integral, since in deforming the contour 
integral onto the real axis, one needs to go around the (Bose) frequency 
$\omega_{n=0}=0$.  Differentiating Eq.~(\ref{omegafreq}) with respect to
$T$ at constant $A$ and $B$, we decompose the entropy into a sum of three
terms,
\begin{eqnarray}
S=-\left.{\partial\Omega\over\partial T}\right|_{\mu,V,A,B}&=&S_f+S_b+S',
\label{ssfsbs}
\end{eqnarray}
where
\begin{eqnarray}
S_f&\equiv&-\sum_{\bf p}\int{d\omega_p\over  \pi} \,{\partial f\over \partial
T}{\rm tr\,}\left( {\rm Im} \Sigma\, {\rm Re} G+{\rm Im}\,
\log(-\gamma_0 G^{-1})\right),\label{sf} \\
S_b&\equiv&-\sum_{{\bf q},l}g_l\pp{d\omega_q\over 2 \pi} \,{\partial
n\over \partial T} \left({\rm Im}\Pi_l\,{\rm Re} D_l+{\rm Im}
\log(-D_l^{-1})\right)
,\label{sb} \\
S'&\equiv&\left.-{\partial (T \Phi)\over\partial T}\right|_{A,B} 
-\sum_{\bf p}\int{d\omega_p\over  \pi}
\,{\partial f\over \partial T} {\rm tr\,} ({\rm Im} G \,{\rm Re} \Sigma)
\nonumber \\
&& -\sum_{{\bf q},l} g_l \pp{d\omega_q\over 2 \pi}\, {\partial n\over \partial
 T}\,{\rm Im} D_l \,{\rm Re} \Pi_l.
\label{sp}
\end{eqnarray}
In this decomposition, the terms $S_f$ and $S_b$ are the contributions
from the electron and photon elementary modes, while $S'$ is a correction
term arising from the interactions.

    We illustrate the derivation in the one--loop approximation, which
corresponds to taking the diagram $\Phi$ depicted in Fig.~1, and the
self-energy diagrams of Fig.~2.  Applying the usual Feynman rules, and
carrying the sum over frequencies, we find
\begin{eqnarray}
T \Phi&= &
{e^2\over16  \pi^3 V}\sum_{{\bf p},{\bf q},l}
g_l \int{d\omega_p d\omega_{p'} d\omega_q} P_l
B_l(\omega_q,q)
\frac{n f
\left(1-f'\right)-\left(1+n\right) \left(1-f\right)
f'}{\omega_p+\omega_q-\omega_{p'}}+T \Phi_{{\rm HF}} \label{phi2}, 
\nonumber  \\ \\
\Sigma(z,p)&= &{e^2\over 4\pi^2 V}
\sum_{{\bf q},l} g_l \int d\omega_q d\omega_{p'} S_l(\omega_{p'},p')
B_l(\omega_q,q)\, \frac{\left(1-f'\right)n+
f'\left(1+n\right)}{z+\omega_q-\omega_{p'}} +\Sigma_{\rm HF}, \label{sigma2}\\
\Pi_l(z,q)&= &{e^2\over 4\pi^2 V}\sum_{{\bf p}}\int d\omega_pd\omega_{p'}
P_l\, \frac{f\left(1-f'\right)-
f\left(1-f'\right)}{z+\omega_p-\omega_{p'}},
\label{pi2}
\end{eqnarray}
where $f\equiv f(\omega_p)$, $f'\equiv f(\omega_{p'})$
and $n\equiv n(\omega_q)$,
$z$ is complex, ${\bf p'}={\bf p}+{\bf q}$, and the
matrix elements $P_l$ and $S_l$ are given by
\begin{eqnarray}
P_L(\omega_p,p;\omega_{p'},p')&\equiv&
{\rm tr\,}\left[\gamma^0 A(\omega_p,p) \gamma^0 A(\omega_{p'},p')\right], \\
P_T(\omega_p,p;\omega_p,p')&\equiv& \frac12{(\delta_{ij}-{\hat{\bf{q}}}_i
{\hat{\bf{q}}}_j)}\,
{\rm tr\,}\left[\gamma^i A(\omega_p,p) \gamma^j A(\omega_{p'},p')\right], \\
S_L(\omega_{p'},p')&\equiv& \gamma^0 A(\omega_{p'},p')\gamma^0, \\
S_T(\omega_{p'},p')&\equiv& \frac12 {(\delta_{ij}-{\hat{\bf{q}}}_i
{\hat{\bf{q}}}_j)}\, \gamma^i A(\omega_{p'},p')\gamma^j.
\end{eqnarray}
The terms $\Phi_{\rm HF}$ and $\Sigma_{\rm HF}$ in Eqs.~(\ref{phi2}) and
(\ref{sigma2}) arise from the static term $1/q^2$ in the photon
propagator of Eq.~(\ref{Dlspec}),
\begin{eqnarray}
T \Phi_{\rm HF}&=& -{e^2 \over 8 \pi^2 V} \sum_{{\bf p},{\bf p'}}\int
{d\omega_p}{d\omega_{p'}}\, {\rm tr\,}\left[\gamma^0 A(\omega_p,p) \gamma^0 
A(\omega_{p'},p')\right]
\frac{f f'}{q^2}, \label{phiHF}\\
\Sigma_{\rm HF}&=& - {e^2 \over 2\pi V} \sum_{{\bf p'}} \int d\omega_{p'}\,
\gamma^0 A(\omega_{p'},p') \gamma^0\, \frac{f'}{q^2} \label{sigmaHF}.
\end{eqnarray}

    In the one-loop approximation $S'$ in fact vanishes.  (A similar result
was observed in the SPA approximation in liquid $^3$He by 
Riedel~\cite{Riedel}.)
To see how the various terms in Eq.~(\ref{sp}) cancel, we first note that the
same combination of matrix elements appears in each term of $S'$.  Since
${\rm Im}[G(\omega_p+i 0^{+})]=-A(\omega_p,p)/2$ and ${\rm Im} [D_l(\omega_p+i
0^{+})]=-B_l(\omega_q,q)/2$,
\begin{eqnarray}
S'&=&-{e^2 \over 8\pi^2 V}\sum_{{\bf p},{\bf p'}} \int d\omega_p d\omega_{p'}\,
{\rm tr\,}\left[\gamma^0 A(\omega_p,p) \gamma^0 A(\omega_{p'},p')\right]
\frac{{\cal S}_{\rm HF}}{q^2}\nonumber \\
&& +
{e^2\over16  \pi^3 V}\sum_{{\bf p},{\bf q},l} g_l
\pp{d\omega_p d\omega_{p'} d\omega_q}\,  P_l(\omega_p,p;\omega_{p'},p') 
B_l(\omega_q,q)\, \frac{{\cal S} }{\omega_p+\omega_q-\omega_{p'}},
\end{eqnarray}
where the ${\cal S}$ and ${\cal S}_{\rm HF}$ denote combinations of
statistical factors and their T derivatives.  Using Eqs. (\ref{phi2}) through
(\ref{sigmaHF}),
symmetrizing the contributions of ${\rm Re}\, \Sigma_{\rm HF}$ and 
${\rm Re}\,\Sigma$ by the transformation $(\omega_p,{\bf p})$
$\leftrightarrow$$(\omega_{p'},{\bf p'}),$  $(\omega_q,{\bf q})$
$\leftrightarrow$$(-\omega_q,-{\bf q})$,
and using the fact that the spectral functions $B_l$ are odd
functions of their arguments, we have
\begin{eqnarray}
{\cal S}_{\rm HF}&=& -{\partial \left\{f f'\right\}/\partial T}+ 
f' {\partial f/\partial T}+ f {\partial f'/\partial T}=0, \label{spHF} \\
{\cal S} &=&
-{\partial \left\{n f (1-f')-(1+n)(1-f) f'\right\}/\partial T}+
(f(1-f')-f'(1-f))\,{\partial n/\partial T}
\nonumber \\
&+& (n(1-f')+(1+n) f')\,{\partial f/\partial T}
+(nf+(1+n)(1-f))\,{\partial f'/\partial T}=0.
\label{sp2}
\end{eqnarray}
Thus the correction term $S'$ vanishes in the one-loop approximation.

    This cancelation takes place only in the lowest order diagram for
$\Phi$.  A general analysis of $\Phi$ diagrams in the context of
non-relativistic normal Fermi liquids\cite{He3} shows that the
contributions to $S'$ come only from those graphs that have at least
two vanishing energy denominators.  These terms correspond graphically
to the set of diagrams that can be cut in three and only three
different pieces when one removes a set of fermion lines.
Generalizing this analysis to the diagrams representing the functional
$\Phi[G,D]$, we see that there is only one possible cut in the diagram
of Fig. 1 and it generates the vanishing denominator
$\omega_q+\omega_p-\omega_{p'}$.  The simplest diagram that contributes to 
$S'$ is shown in Fig.~3, where we have drawn a set of cuts that give two
vanishing energy denominators.

\section{Interpretation of the entropy formula}

\subsection{Exchange and correlation entropy}

    When the two electron lines in Fig. 1 are replaced by bare
propagators and the photon line is replaced by a dressed propagator in
the RPA approximation, i.e., with the lowest order bubble diagram
insertions, we recover the set of diagrams that contribute to the
exchange and correlation terms considered by Akhiezer and
Peletminski\v\i ~\cite{AkhiPele60}.  It is instructive to see how the
expression of the entropy that we have derived, Eqs.~(\ref{sf}),~(\ref{sb})
and (\ref{sp}), generates these exchange and correlation terms correctly
when expanded in the first few powers of the coupling constant $e$.
We start with the exchange entropy~\cite{AkhiPele60},
\begin{eqnarray}
S_s&\equiv&{1\over 2}{\partial \over \partial T}\left(T\, {\rm Tr\,}\Pi^{(2)} 
D_0\right) =\sum_{\bf q}\pp{d\omega_q\over 2\pi} {\partial \over \partial T}
\left(n\, {\rm Im}\,\Pi^{(2)}\,D_0\right),
\end{eqnarray}
where $\Pi^{(2)}$ is the photon self--energy of Fig. 2 with two bare
electron propagators.  [In this section, we do not  write the sum over 
polarization states explicitely.]  Expanding Eqs.~(\ref{sf}) and (\ref{sb})
to first order in $e^2$, we find
\begin{eqnarray}
S^{(2)}=\sum_{\bf p}\int {d\omega_p\over \pi}\, {\partial f\over \partial T} 
{\rm tr\,}{\rm Im} G_0\,{\rm Re}\, \Sigma[D_0] +\sum_{\bf q}\pp 
{d\omega_q\over 2\pi}\, {\partial n\over \partial T}
{\rm Im}\,D_0\,{\rm Re}\,\Pi^{(2)}.
\label{s2}
\end{eqnarray}
Then expanding Eq.~(\ref{sp}) for $S'=0$ to order $e^2$, we recognize that
the right side of Eq.~(\ref{s2}) is $S^{(2)}=-{\partial (T \Phi^{(2)})/ 
\partial T}$, which from the identity $\Phi^{(2)}=(-1/2) {\rm Tr\,}
\Pi^{(2)} \, D_0$ becomes $S^{(2)}=(1/2){\partial (T\,{\rm Tr\,}
\Pi^{(2)}\,D_0)/\partial T}=S_s$.

    The sum of the exchange and the correlation terms that contribute to the
entropy is~\cite{AkhiPele60}
\begin{eqnarray}
S_c+S_s &\equiv& -{1\over 2}{\partial \over \partial T}\left(T
\,{\rm Tr\,}\log(-1+D_0 \Pi^{(2)})\right)\nonumber \\
&=&-\sum_{\bf q} \pp {d\omega_q\over 2\pi}
\left({\partial n\over \partial T}\,
{\rm Im} \left[\log(-1+D_0 \Pi^{(2)})\right]
-n\, {\rm Im}\left[ D\,{\partial\Pi\over \partial T}^{(2)}\right]\right),
\label{sc}
\end{eqnarray}
where $D$ is now the propagator in the RPA approximation,
$D^{-1}=D_0^{-1}-\Pi^{(2)}$.  Using arguments similar to those we used to
derive the identity $S'=0$, we can decompose the last term on the right side
into the following terms:
\begin{eqnarray}
\sum_{\bf q} \pp {d\omega_q\over 2\pi}\, n\,{\rm Im}\left[ D\,{\partial\Pi\over
\partial T}^{(2)}\right]&=&\sum_{\bf p}\int{d\omega_p\over \pi}\,{\partial f 
\over \partial T}
{\rm Im}\,G_0\, {\rm Re}\,\Sigma[D]\nonumber \\
&&-\sum_{\bf q}\pp{d\omega_q\over 2\pi}
\,{\partial n \over \partial T}\, {\rm Im}\,\Pi^{(2)}\,{\rm Re}\,D.
\end{eqnarray}
The two terms on the right side correspond respectively to $S_f$ expanded
to linear order in $\Sigma[D]$, and to the first term in $S_b$.  Adding the
first term on the right side of Eq.~(\ref{sc}), we have therefore shown that 
our expression for $S=S_f+S_b$ gives the correct correlation term in the 
entropy when one uses bare electrons and the RPA photon propagator in
Eqs.~(\ref{sf})-(\ref{sp}).

\subsection{Decomposition in elementary excitation modes}

    The formulae we obtained for the entropy terms $S_f$, Eq.~(\ref{sf}), and
$S_b$, Eq.~(\ref{sb}), allow us to perform a spectral analysis by decomposing
$S_f$ and $S_b$ into integrals over the elementary excitations of the matter
field and the electromagnetic field.  We closely follow the derivation that
was carried out by Carneiro and Pethick for liquid $^3$He~\cite{He3}.  The key
is to transform the temperature derivatives of the statistical factors $f$ and
$n$ into the following expressions
\begin{eqnarray}
{\partial f\over \partial T}&=&-{\partial \sigma_f\over\partial \omega_p}
(\omega_p), \label{dft:sigmaf}\\
{\partial n\over \partial T}&=&-{\partial \sigma_n\over\partial \omega_q}
(\omega_q), \label{dnt:sigman}
\end{eqnarray}
where again $\sigma_f\equiv -f \log f-(1-f)\log(1-f)$ and $\sigma_n\equiv
-n\log n +(1+n)\log(1+n)$ are the entropy contributions from an electron
mode of energy $\omega_p$ and from a photon mode of frequency
$\omega_q$. Integrating Eqs.~(\ref{sf}) and~(\ref{sb}) by parts, we see that 
$S_f$ and $S_b$ obey the spectral representations
\begin{eqnarray}
S_f&=&\sum_{\bf p}\int {d\omega_p\over 2\pi}\,\sigma_f(\omega_p)\,
A_s(\omega_p,p),
\label{sfspec}\\
S_b&=&\sum_{\bf q}\pp_0^\infty {d\omega_q\over
2\pi}\,\sigma_b(\omega_q) B_s(\omega_q,q),
\label{sbspec}
\end{eqnarray}
where $A_s$ and $B_s$ are defined by
\begin{eqnarray}
A_s(\omega_p,p)&=&{\rm tr\,}{\partial\over\partial \omega_p}\left\{{\rm Re} 
G\,\Gamma + 2 {\rm Im}\,\log (-\gamma_0 G)\right\},
\label{as}\\
B_s(\omega_q,q)&=&\sum_l g_l\,{\partial\over\partial \omega_q}\left\{{\rm Re} 
D_l\,L_l + 2 {\rm Im} \log (-D_l)\right\},
\label{bs}
\end{eqnarray}
while $\Gamma(\omega_p,p)\equiv-2\,{\rm Im}\,\Sigma(\omega_p+i 0^{+},p)$ and
$L_l(\omega_q,q)\equiv-2\,{\rm Im}\,\Pi_l(\omega_q+i 0^{+},q)$ are the 
imaginary parts of the electron and photon self--energies.  In deriving
Eq.~(\ref{sbspec}), we have used the fact that the boson propagator $D$
is an even function of its arguments to reduce the domain of
integration in Eq. ~(\ref{sbspec}) to positive frequencies $\omega_q$.

    The sum over spinor indices (the trace ${\rm tr\,}$) in Eq.~(\ref{as}) 
can easily be decomposed into two contributions, one from electron states 
with a chirality equal to their helicity (subscript $+$) and one from states 
with opposite helicities and chiralities (subscript $-$), by writing
\begin{eqnarray}
G&=&{\gamma_0-{\bf \gamma}\cdot {\hat {\bf p}}\over 2} \,G_+ +
{\gamma_0+{\bf \gamma}\cdot {\hat {\bf p}}\over 2} \,G_- \nonumber \\
\Sigma&=&{\gamma_0-{\bf \gamma}\cdot {\hat {\bf p}}\over 2} \,\Sigma_- +
 {\gamma_0+{\bf \gamma}\cdot {\hat {\bf p}}\over 2} \,\Sigma_+,
\end{eqnarray}
which gives
\begin{eqnarray}
{\rm tr\,} {\rm Re}\, G\, \Gamma = 2\left({\rm Re}\, G_+\,\Sigma_+ +{\rm Re}\,
G_-\,\Sigma_-\right),\label{decomp}
\end{eqnarray}
where the factor $2$ is from the spin sum.  Hence, $A_s=2\sum_\pm A_{s\pm}$.

    We now turn to the entropy spectral functions $A_s$ and $B_s$.  In the
one-loop approximation of Fig. 1, the self-energies $\Sigma$ and $\Pi$ of
Eqs.~(\ref{sigma2}) and (\ref{pi2}) depend on the fully dressed spectral 
functions $A$ and $B$.  These functions depend in turn on the self-energies 
$\Sigma$ and $\Pi$ through Dyson's equations, Eq.~(\ref{Dyson}).  The problem 
of determining $A$ and $B$, and the functions $A_s$ and $B_s$, is therefore a 
self-consistent one.  However, we can make general statements about their 
structure on the basis of the following arguments.  If the system develops 
well-defined excitation modes, we expect the functions $A$ and $B$ to consist 
of narrow peaks at the locations of the quasiparticles and collective modes, 
as well as wide bands of continuum states.  The continuum states that 
contribute to the spectral function $A$ are composed of particle-photon states,
while those contributing to $B$ are particle-hole and particle-antiparticle 
states.  In lowest order of perturbation theory, the continua extend over 
frequency ranges $-p<\omega_p<p$ and $-q<\omega_q<q$, 
respectively~\cite{plasmons,fermions}.  We expect
that in the present self-consistent problem, interactions only slightly modify
these frequency ranges.  The quasiparticle modes are at frequencies well
outside the spectrum of continuum states, $|\omega_p|>p$ and $\omega_q>q$.

    The structure of the entropy spectral functions $A_s$ and $B_s$ is
qualitatively similar to that of $A$ and $B$.  The photon spectral function
$B_s$, as we see from its definition, Eq.~(\ref{bs}), has a support over the
frequency range of the continuum states contributing to $B$.  It also contains
a contribution from the oscillation modes of the electromagnetic field, i.e.,
the longitudinal and transverse plasmon modes, with frequencies $\omega_L(q)$
and $\omega_T(q)$ and momentum ${\bf q}$, where $\omega_{L,T}(q)>q$.  As we
show, the spectral function $B_s$ takes a simple form in the vicinity of these
frequencies
\begin{eqnarray}
B_{s,l}(\omega_q,q)&\simeq& \frac{(Z_{l} L_{l})^3/2}
{((\omega_q-\omega_{l}(q))^2+(Z_{l} L_{l}/2)^2)^2},
\label{Bspole}
\end{eqnarray}
where $Z_l\equiv{\partial {\rm Re} D_{l}^{-1}/\partial \omega_q}$, $Z_L\simeq
\omega_L/(2 q^2)$ and $Z_T\simeq 1/(2\omega_T)$.  Evaluating the logarithmic 
term in Eq.~(\ref{bs}), we have for each polarization state $l$,
\begin{eqnarray}
2 {\rm Im} \log(-D_{l})&=& 2\pi \Theta(-{\rm Re} D_{l}^{-1})-2 \arctan
{L_{l} \over 2 {\rm Re} D_{l}^{-1}},
\end{eqnarray}
where the values of the arc tangent are in the range $[-\pi/2,\pi/2]$.
The function $L_{l}/(2\omega_q)$ is the interaction rate of an excitation
of frequency $\omega_q$.  In the vicinity of the modes, $\omega_q
\simeq \omega_{L,T}(q)$, $L_l/2 \omega_q$ is perturbatively smaller than the 
modes frequencies $\omega_{L,T}(q)$~\cite{plasmondamping}, and varies 
slowly around $\omega_{L,T}(q)$. Thus, with 
${\rm Re} D_l={\rm Re} D^{-1}_l/(({\rm Re} D_l^{-1})^2+ (L_l/2)^2)$, and
${\rm Re} D_l^{-1} \simeq (\omega_q-\omega_l(q))/Z_l$, 
we can evaluate the right side of Eq.~(\ref{bs}) 
keeping $L_l$ constant, and derive Eq.~(\ref{Bspole}).

    The spectral functions $B_{s,l}$ are therefore sharply peaked around the
mode frequency $\omega_{l}$.  Comparing to the photon spectral densities $B_l$,
which have a Lorentzian form close to the poles, $B_{l}\simeq (Z_{l} L_{l})/
((\omega_q-\omega_{l})^2+(Z_{l} L_{l}/2)^2)$, we see that $B_{s,l}$ have 
a stronger peak and smaller wings.  
The electron spectral function, $A_s=2\sum_\pm A_{s\pm}$, has a
structure similar to that of $B_s$:
continuum states contribute at small energies $|\omega_p|\leq p$,
while $A_{s\pm}$ exhibit a sharp variation close to the quasiparticle energies
$\omega_{0\pm}$,
\begin{eqnarray}
A_{s\pm}(\omega_p,p)&\simeq& \frac{(Z_{\pm} \Gamma_{\pm})^3/2}
{((\omega_p-\omega_{0\pm})^2+(Z_{\pm} \Gamma_{\pm}/2)^2)^2},
\label{aspole}
\end{eqnarray}
where $Z_{\pm}$ is the quasiparticle mode residue.  In deriving
Eq.~(\ref{aspole}), we have neglected the variation of the (one-loop
order) interaction rate $\Gamma_{\pm}$, which for energies $\omega_p$ in
the vicinity of $\omega_{0\pm}$ is $\sim {\cal O}(e^2 T
\log({q_D}/|\omega_p-\omega_{0\pm}|))$, where $q_D\sim e T$ is the Debye
momentum\cite{irfermion}.

    Both the electron and the photon excitation spectra exhibit collective
modes at small momenta.  However, the low energy modes contribute to the total
entropy in very different orders in the coupling constant.  The contribution
from long wavelength modes of the electromagnetic field has been calculated by
Akhiezer and Peletmniski\v\i ~\cite{AkhiPele60}; the result is that the sum of
plasmon oscillations and continuum modes contributes to the correlation term
in the entropy, of order $e^3 T^3$.  We can easily understand this order of
magnitude by noting that the contribution to the entropy of all states of
energies of order of the plasmon frequency $\omega_{L,T}(0)\sim e T$ is
$\sim \sum_{\bf q}\propto q_c^3$, where the momentum cutoff is $q_c\sim e T$.
The small energy modes in the electron spectrum give a smaller contribution,
of order $e^5 T^3$.  To see this, we expand to lowest order the difference of
$S_f$ from its expression for a non-interacting gas, and concentrate on the
phase space of small energy modes, $\omega_p \sim e T$,
\begin{eqnarray}
\Delta S_f \simeq 2
\sum_{\pm}\sum_{\bf p} \int {d\omega_p \over \pi} {\partial f\over \partial T}
A_{0\pm}(\omega_p,p)\,{\rm Re}\,\Sigma^{(2)}_{\pm}(\omega_p,p).
\end{eqnarray}
For small energies, $\omega_p\sim eT$, ${\partial f/ \partial T}$ is $\sim 
\omega_p/(4 T^2)$, and $A_{0\pm}(\omega_p,p)\,{\rm Re}\,\Sigma_{\pm}^{(2)}
(\omega_p,p)$ is $\simeq\pm 2\pi m^2_f\,\delta(\omega_p\mp p)/p$, where $m_f$ 
is the gap at zero momentum.  Thus, $\Delta S_f \sim m_f^2/T^2\sum_{\bf p}
\sim e^2 p_c^3 \sim e^5 T^3$, for a cutoff momentum $p_c\sim e T$. The reason 
why $\Delta S_f$ is higher by two orders in the coupling constant than the
contribution to the total entropy from long wavelength modes of the
electromagnetic field is the following. An expansion in power series
of $e$ of the logarithm term in $B_s$, Eq.~(\ref{bs}), 
leads to infrared divergences term by term~\cite{AkhiPele60}; hence, to
evaluate the correlation term in the entropy, one must include the
logarithm as it stands. However, an expansion of the logarithm term in
$A_s$, Eq.~(\ref{as}), is divergence-free and results in a prefactor
$m^2_f/T^2 \sim e^2$.

    Although the one-loop approximation is mathematically tractable,
it has important limitations.  To take into account correctly the low
energy part of the electron spectrum, which enters only in order $e^5
T^3$, one needs to consider $\Phi$-diagrams of higher orders, such as
the one depicted in Fig. 3, contributing at least in order $e^4$.
Also, the diagram of Fig. 1 does not always provide the correct width
for the entropy spectral functions evaluated close to the
quasiparticle modes.  For instance, the lifetime of a photon mode of
energy $\sim T$ is limited by Compton scattering and inverse pair
annihilation.  The self-consistent approximation of Fig. 1 includes
only the direct terms of these processes, as can be seen by
considering an electron self-energy insertion in one of the
fermion lines of the bubble diagram shown in Fig. 2 and then taking
the imaginary part of the corresponding photon self-energy. 
The cross terms arise from the imaginary part of
the photon self-energy obtained by removing one photon line in Fig.~3.
Finally, in calculating the widths of low energy modes, one also needs
to include vertex corrections, as described by the ``Hard Thermal
Loop'' perturbation scheme, see Ref.~\cite{HTL}.

    We conclude by stressing the fact that our analysis of the entropy
provides a framework for studying effects of the infrared structure in
the fermion propagator on thermodynamical quantities.  Kim et al. have
proposed, in the context of the fractional quantum Hall
effect~\cite{Cvgauge}, that one should be careful in
calculating thermodynamical quantities by summing over fermion degrees
of freedom, since by doing so, one may encounter infrared divergent terms. 
Kim et al. attribute this problem to the fact that the composite fermion 
propagator is not a gauge invariant quantity and suggest that the 
thermodynamical quantities should be calculated instead by  summing over boson 
degrees of freedom only. In the
present analysis, the dangerous term is the first component of $S_f$,
$\sim \sum_{\bf p} \int (d\omega_p/2\pi){\partial f/\partial T}\,{\rm
Re}\,G \Gamma$, with $\Gamma \propto e^2 T \log(q_D/|\omega_p-p|)$,
and it does not lead to a divergence, as the principal part $\pp d x
\log|x|/x$ vanishes.  This term is actually finite and is part of the exchange
entropy term, of order $e^2$.  On the other hand, for energies
$\omega_p$ in the vicinity of $p$, both the spectral densities
$A\simeq \Gamma/((\omega_p-p)^2+(\Gamma/2)^2)$ and $A_s\simeq
\Gamma^3/2/((\omega_p-p)^2+(\Gamma/2)^2)^2$ vanish as $\sim
1/\Gamma\sim\log^{-1}|\omega_p-p|$, as $\omega_p \sim p$. 
This logarithmic behavior is symptomatic of a breakdown of perturbation 
theory. In a future publication we will examine corrections in the free 
energy similar to those considered in the calculation of the fermion 
lifetime in a hot plasma \cite{irfermion}.

\section*{Acknowledgments}

    We warmly thank Dr.~B.~Bl\"attel of BMW, Munich, for his participation in
the early stages of this work.  This research was supported in part by NSF
Grants PHY94-21309 and PHY89-21025.

\newpage
\centerline{\bf FIGURE CAPTIONS}

\vspace*{0.5cm}
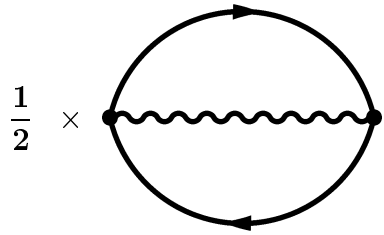
\noindent Figure 1:
$\Phi$ in the one-loop approximation.

\vspace*{0.5cm}
\noindent Figure 2:
a) One-loop electron self-energy. b) One-loop photon self-energy.

\vspace*{0.5cm}
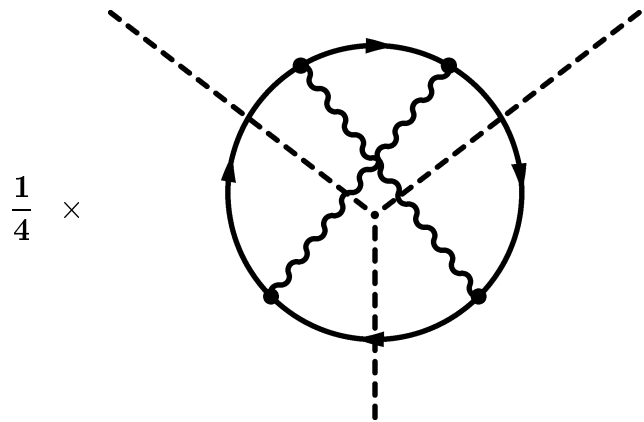
\noindent Figure 3:
$\Phi$-diagram to explicit order $e^4$.  The dashed lines illustrate the
cuts that give rise to a non-vanishing term in $S'$.


\begin{thebibliography}{99}

    \bibitem{BaymKada61} G. Baym and L.P.  Kadanoff, {\it Phys.  Rev.}
{\bf 124}:287 (1961).

    \bibitem{Phiconser2} G. Baym, {\it Phys.  Rev.}  {\bf 127}:1391 (1962).

\bibitem{CollPer75}
P. Carruthers, {\it Collective phenomena} {\bf 1}:147 (1973);
J. C. Collins and M. J. Perry, {\it Phys. Rev. Lett.} {\bf 34}:1353 (1975);
G. Baym and S. A. Chin, {\it Phys. Lett.} {\bf 62B}:241 (1976);
M. Kislinger and P. Morley, {\it Phys. Lett.} {\bf 67B}:371 (1977).

    \bibitem{QM96} See the proceedings of the ongoing Quark Matter
conferences, e.g., {\it Nucl.  Phys.  {\bf A610}} (1996).

    \bibitem{HalpLeeRead95} B.I.  Halperin, P.A.  Lee, and N. Read, {\it Phys.
Rev.}  {\bf B47}:7312 (1993); L.B.  Ioffe and A.I.  Larkin, {\it Phys.  Rev.}
{\bf B39}:8988 (1989).

    \bibitem{AkhiPele60} I. A. Akhiezer and S. V. Peletminski\v\i, {\it Sov.
Phys. JETP} {\bf 11}:1316 (1960).

    \bibitem{FreeMcLe77} B. A. Freedman and L. D. McLerran, {\it Phys.  Rev.}
{\bf D16}:1130, 1147, 1169 (1977).

    \bibitem{thermopot}
V. Baluni, {\it Phys.  Rev.} {\bf D17}:2092 (1978); C. Coriano and
R. R. Parwani, {\it Phys. Rev. Lett.} {\bf 73}:2398 (1994),
{\it Nucl. Phys.} {\bf B434}:56 (1995);
P. Arnold and C. Zhai, {\it Phys. Rev.} {\bf D50}:7603 (1994);
R. R. Parwani, {\it Phys. Lett.} {\bf B334}:420 (1994);
B. Kastening and C. Zhai, {\it Phys. Rev.} {\bf D52}:7232 (1995);
E. Braaten and A. Nieto, {\it Phys. Rev.} {\bf D53}:3421 (1996).

    \bibitem{plasmons} V.P.  Silin, {\it Sov.  Phys.  JETP }{\bf 11}:1136
(1960); H.A.~Weldon, {\it Phys.  Rev.}  {\bf D26}:1394 (1982).

    \bibitem{fermions} V.V.~Klimov, {\it Sov.  J. Nucl.  Phys.} {\bf 33}:934
(1981);
H.A.~Weldon, {\it Phys.  Rev.}  {\bf D26}:2789 (1982);
G. Baym, J.-P.  Blaizot, and B. Svetitsky, {\it Phys.  Rev.} {\bf D46}:4043
(1992); J.P.  Blaizot and J.-Y.
Ollitrault, {\it Phys.  Rev.}  {\bf D48}:1390 (1993).

    \bibitem{reuter} G.E.H.  Reuter and E.H.  Sondheimer, {\it Proc. Roy. Soc.}
{\bf A195}: 336 (1948).

    \bibitem{BaymMoni} G. Baym, H. Monien, C. J. Pethick, and D. G. Ravenhall,
{\it Phys.  Rev.  Lett.}:1867  {\bf 64} (1990).



    \bibitem{BraaThom} E. Braaten and M.H. Thoma, {\it Phys.  Rev.}
{\bf D44}:1298,2625 (1991).

    \bibitem{HTL} E. Braaten and R.D.  Pisarski, {\it Phys.  Rev.  Lett.}
 {\bf 64}:1338 (1990); {\it Phys.  Rev.} {\bf D42}:2156 (1990);
{\it Nucl.  Phys. } {\bf B337}:569 (1990).


    \bibitem{kinetic} J.-P.  Blaizot and E. Iancu, {\it Phys.  Rev.  Lett.}
{\bf 70}:3376 (1993); {\it Nucl.  Phys.}  {\bf B390}:589 (1993).

    \bibitem{irfermion} J.-P.  Blaizot and E. Iancu, {\it Phys.  Rev.  Lett.}
{\bf 76}:3080 (1996); {\it Phys.  Rev.} {\bf D55}:973 (1997).

\bibitem{MuccLee}
P.A. Lee, E.R. Mucciolo, and H. Smith, {\it Phys. Rev.} {\bf B54}:8782 (1996).


    \bibitem{Cvgauge} Y. B. Kim, P. A. Lee, X.-G.  Wen, and P. C. E. Stamp,
{\it Phys.  Rev.} {\bf B51}:10779 (1995); Y. B. Kim, and P. A. Lee,
{\it Phys.  Rev.} {\bf B54}:2715 (1996).

    \bibitem{screepert} F. Karsch, A. Patkos and P. Petreczky,
{\it Phys.  Lett.} {\bf B401}:69 (1997); A. Peshier, B. K\"ampfer,
O.P.  Pavlenko and G. Soff, hep-ph/9801344.

    \bibitem{Phiconser1} J. M. Luttinger and J. C. Ward, {\it Phys.  Rev.}
{\bf 118}:1417 (1960).

    \bibitem{He3} G. M. Carneiro and C. J. Pethick, {\it Phys.  Rev.}
{\bf B7}:304 (1973); {\it Phys.  Rev.} {\bf B11}:1106 (1975), and references
therein.

    \bibitem{Riedel} E. Riedel, {\it Z.  Phys.}  {\bf 210}:403 (1968).

    \bibitem{Gotze} W. G\"otze and H. Wagner, {\it Physica} {\bf 31}:475
(1965); G. Baym and G. Grinstein, {\it Phys.  Rev.} {\bf Dl5}:2897 (l977);
A. Griffin, {\it Phys. Rev.} {\bf B53}:9341 (1996).

    \bibitem{remark} 
The limit of ${\rm Tr\,}\log(-D^{-1})$ as $\lambda \to 0$ keeps only
transverse polarization states.  In Coulomb gauge, $D_L^{-1}\equiv
q^2-\Pi_L$ and $D_T^{-1}\equiv \omega_n^2-q^2-\Pi_T$.  The trace
${\rm tr\,}\equiv\sum_{{\bf q},n}{\rm tr\,}$ contains a trace over momenta,
frequencies and Lorentz indices.  The trace over Lorentz indices gives
${\rm tr\,} \log (-D^{-1})=$$\log\, {\rm det}(-D^{-1})$$=\log( (q^2-\Pi_L)
(\omega^2-q^2-\Pi_T)^2)$.  In the limit $\lambda\to 0$,
$\Pi_L=\Pi_T=0$, and the frequency sum over the longitudinal term
vanishes, ${\rm tr\,} \log\,q^2= \sum_{\bf q}\log\,q^2\sum_n=0$.  Thus, only
the transverse modes contribute, with a factor 2 for the two
polarization states of real photons.


    \bibitem{BaymKada62} L.P.  Kadanoff and G. Baym, {\it Quantum Statistical
Mechanics} (W.A.  Benjamin, New York, 1962).

    \bibitem{plasmondamping}
R.D. Pisarski, {\it Phys. Rev.} {\bf D47}:5589 (1993); see also Refs. 14.

\end{thebibliography}
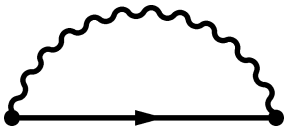
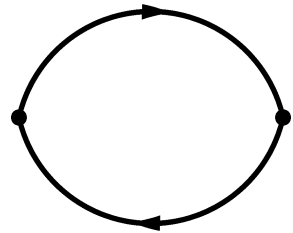
\end{document}